\documentclass[10pt,twocolumn,letterpaper]{article}

\usepackage{wacv}
\usepackage{times}
\usepackage{epsfig}
\usepackage{graphicx}
\usepackage{amsmath}
\usepackage{amssymb}
\usepackage{bm}
\usepackage{subfig}

\def\wacvPaperID{346} % Enter the WACV Paper ID here

\wacvfinalcopy % *** Uncomment this line for the final submission

\ifwacvfinal
\def\assignedStartPage{1} % *** Enter the assigned starting page number (instead of 9876)
\fi

\ifwacvfinal
\usepackage[breaklinks=true,bookmarks=false]{hyperref}
\else
\usepackage[pagebackref=true,breaklinks=true,colorlinks,bookmarks=false]{hyperref}
\fi

\ifwacvfinal
\else
\fi

\begin{document}

%%%%%%%%% TITLE
\title{Ensembling Low Precision Models for Binary Biomedical Image Segmentation}

\author{Tianyu Ma\\
Cornell University\\
{\tt\small tm478@cornell.edu}
\and
Hang Zhang\\
Cornell University\\
{\tt\small hz459@cornell.edu}

\and
Hanley Ong\\
Weill Cornell Medical College\\
{\tt\small hao2007@med.cornell.edu}

\and
Amar Vora\\
Weill Cornell Medical College\\
{\tt\small apv7002@med.cornell.edu}

\and
Thanh D. Nguyen\\
Cornell University\\
{\tt\small tdn2001@med.cornell.edu}

\and
Ajay Gupta\\
Weill Cornell Medical College\\
{\tt\small ajg9004@med.cornell.edu}

\and
Yi Wang\\
Cornell University\\
{\tt\small yw233@cornell.edu}

\and
Mert R. Sabuncu\\
Cornell University\\
{\tt\small msabuncu@cornell.edu}
}

\maketitle
%\thispagestyle{empty}

%%%%%%%%% ABSTRACT
\begin{abstract}
Segmentation of anatomical regions of interest such as vessels or small lesions in medical images is still a difficult problem that is often tackled with manual input by an expert. 
One of the major challenges for this task is that the appearance of foreground (positive) regions can be similar to background (negative) regions.
%As a result, segmentation algorithms are often confused between foreground and background, which leads to either under or over segmentation. 
As a result, many automatic segmentation algorithms tend to exhibit asymmetric errors, typically producing more false positives than false negatives. 
In this paper, we aim to leverage this asymmetry and train a diverse ensemble of models with very high recall, while sacrificing their precision.  
Our core idea is straightforward:
A diverse ensemble of low precision and high recall models are likely to make different false positive errors (classifying background as foreground in different parts of the image), but the true positives will tend to be consistent. 
Thus, in aggregate the false positive errors will cancel out, yielding high performance for the ensemble.
%An ensemble of these weak but diverse models can lead to better performance.
Our strategy is general and can be applied with any segmentation model.
In three different applications (carotid artery segmentation in a neck CT angiography, myocardium segmentation in a cardiovascular MRI and multiple sclerosis lesion segmentation in a brain MRI), we show how the proposed approach can significantly boost the performance of a baseline segmentation method.
\end{abstract}

\begin{figure}[t]
\begin{center}
\includegraphics[width=0.9\linewidth]{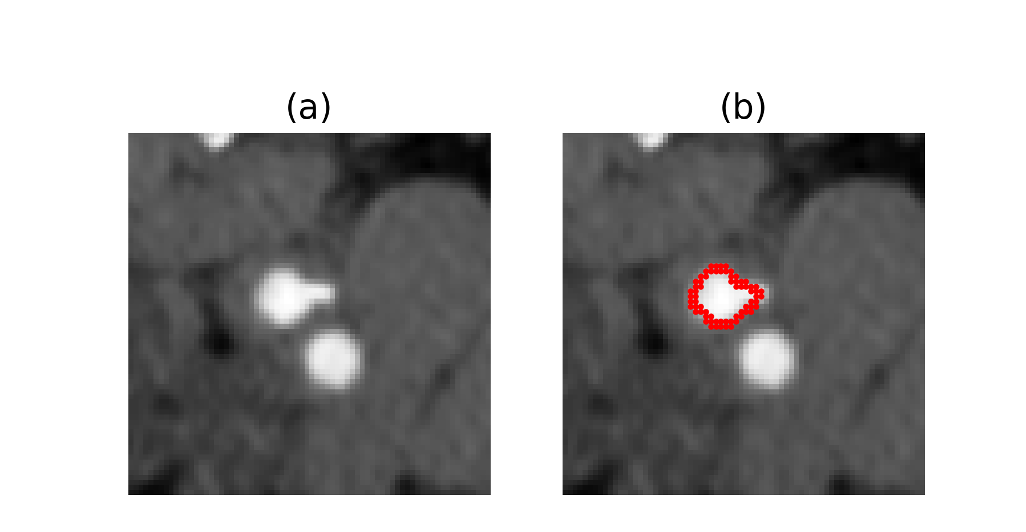}
\end{center}
\caption{(a) An example CTA image with both internal and external carotid artery. (b) The image and overlaid manual ground-truth of the internal carotid artery}
\label{intro}
\end{figure}

%%%%%%%%% BODY TEXT
\section{Introduction}

Deep learning techniques, such as the U-Net~\cite{unet}, produce most state-of-the-art biomedical image segmentation tools. However, delineating a relatively small region of interest in a biomedical image (such as a small vessel or lesion) is a challenging problem that current segmentation algorithms still struggle with. 
\textcolor{black}{In these applications, one of the main problems is that there are different anatomic structures present in the images with similar shapes and intensity values to the foreground structure(s), making it difficult to distinguish them from each other~\cite{automate}. Those intrinsically hard segmentation tasks are considered challenging even for human experts.  
One such task is the localization of the \textit{internal} carotid artery in computed tomography angiography (CTA) scans of the neck. As shown in Figure~\ref{intro},
there are no features that separate the internal and external carotid arteries in the CTA appearance other than their relative position\cite{park2005connectivity}.}
Learning these features, particularly with limited data, can be challenging for convolutional neural networks.

In this paper, we propose an easy-to-use novel ensemble learning strategy to deal with the challenges we mention above. Ensemble learning is an effective general-purpose machine learning technique that combines predictions of individual models to achieve better performance. In deep learning, it often involves ensembling of several neural networks trained separately with random initialization~\cite{ensemble,retina}. 
It has also been used to calibrate the predictive confidence in classification models, particularly when the individual model is over-confident as in the case of modern deep learning models~\cite{uncertainty}. In problem settings where it is more likely to make low precision predictions such as nodule detection, ensemble learning has been used to reduce the false positive rate~\cite{pulmonary,logicAND,lung}. 

Ensemble learning has previously been used for binary segmentation, where multiple probabilistic models are combined, for example by weighted averaging, and the final binary output is computed by thresholding the weighted average at 0.5~\cite{threshold,wang2015hierarchical}. 
\textcolor{black}{The performance of a regular ensemble model depends on both the accuracy of the individual segmenters and the diversity among them~\cite{krogh1995neural,zhang2019confidence}. Previous works using ensembling for image segmentation mainly focus on having diverse base segmenters while maintaining individual segmenter accuracy~\cite{kamnitsas2017ensembles,li2018fully}. In this conventional ensemble segmentation framework, the diverse errors are often distributed between foreground and background pixels.
%while the focus is on keeping each model's segmentation accuracy to be high.
% remove false positives when foreground structures are hard to distinguish from the background. For example, when most models in the ensemble mistakenly classify both internal and external carotid artery as the internal carotid artery due to their visual closeness, a simple ensemble of these models will not fix the problem.
} 

In this paper, we present a different take on ensembling for binary image segmentation.
Our approach is to create an ensemble of models that each exhibit low precision (specificity) but very high recall (sensitivity). These individual models are also likely to have relatively low accuracy due to over-segmentation. 
We achieve this by altering widely used loss functions in a manner that tolerates false positives (FP) while keeping the false negative (FN) rate very low. 
\textcolor{black}{During training of models in an ensemble, the relative weights of FP vs. FN predictions in each model are also selected randomly to promote diversity among them.}   
In this way, each model will produce segmentation results that largely cover the foreground pixels, while possibly making different mistakes in background regions. These weak segmenters have high agreement on foreground pixels and low agreement on the predictions for background pixels. Compared to other ensembling strategies, our method thus does not focus on getting accurate base segmenters, but rather diverse models with high recall.  
%We measure the model diversity using pairwise dice scores among all the models. 
We use two popular and easy-to-implement strategies to create the ensemble:
bagging and random initialization. Each model is trained using a random subset of data and all parameters are randomly initialized to maximize the model diversity. 

The proposed approach is general and can be used in a wide range of segmentation applications with a variety of models.
We present three different experiments.
In the first one, we consider the challenging problem of segmenting the internal carotid artery in a neck CTA scan.
The second experiment deals with myocardium segmentation in a cardiovascular MRI. In the third experiment, we test our method on multiple sclerosis lesion segmentation in a brain MRI. 
Our results demonstrate that the proposed ensembling technique can substantially boost the performance of a baseline segmentation method in all of our experimental settings. 

%-------------------------------------------------------------------------
\section{Methods}
% \textcolor{black}{Inherent ambiguities and similar appearance of structures is a major challenge in medical images segmentation \cite{phiseg,probabilistic}. Ensemble learning can alleviate this problem when the errors made by different models are as independent as possible so that they can be canceled during the aggregation of the predictions. 
% However, when learning with data that has a lot of intrinsic ambiguous structures, all models tend to make highly correlated errors e.g. confusion caused by hyper-intensities, and therefore making the ensembling less effective. 
% Thus, to fully utilize the benefits of ensemble learning, we need an alternative ensembling approach that promotes uncorrelated errors.}   

Our proposed method uses weak segmenters with low accuracy and specificity to collectively make predictions. Although individual models are prone to over segment the image, they are capable of capturing most of the true positive pixels. During the final prediction, an almost unanimous agreement is required to classify a pixel as the foreground in order to eliminate the false positive predictions made by each model. The key is for all models in the ensemble to have a high amount of overlap in the true positive parts and low overlap in the false positive predictions. \textcolor{black}{We can achieve this by modifying the loss function to put more weight on false negative than false positive predictions, and use a random weight for each model in the ensemble. We will show in section 2.3 that this simple modification can encourage diverse false positive errors.}   

Our ensemble approach is different from existing ensemble methods, which usually combine several high accuracy models and use majority voting for the final prediction.    

\subsection{Supervised Learning Based Segmentation}
For a binary image segmentation problem, conditioned on an observed (vectorized) image $\bm{x} \in\mathbb{R}^N$ with $N$ pixels (or voxels), the objective is to learn the true posterior distribution $p(\bm{y}|\bm{x})$ where $\bm{y} \in \{0,1\}^N$; and $0$ and $1$ stand for background or foreground, respectively. 

Given some training data, $\{\bm{x}_i, \bm{y}_i\}$, supervised deep learning techniques attempt to capture $p(\bm{y}|\bm{x})$ with a neural network (e.g. a U-Net architecture) that is parameterized with $\theta$ and computes a pixel-wise probabilistic (e.g., soft-max) output $\bm{f}(\bm{x}, \theta)$, which can be considered to approximate the true posterior, of say, each pixel being foreground. 
So $f^j(\bm{x}, \theta)$ models $p(\bm{y}^j = 1|\bm{x})$, where the superscript indicates the $j$'th pixel.
%We will use $\bmf(\bm{x}, \theta)$ to shorthand 
Loss functions widely used to train these neural networks include Dice, cross-entropy, and their variants. 
The probabilistic Dice loss quantifies the overlap of foreground pixels:
\begin{equation}
    L_{Dice}(\bm{y},\bm{f}(\bm{x}, \theta))=1 - \frac{2\sum_{j=1}^N \bm{y}^j\bm{f}^j}{\sum_{j=1}^N \bm{y}^j + \sum_{j=1}^N \bm{f}^j} \label{eq:Dice}
\end{equation}
Cross-entropy loss, on the other hand, is defined as:
\begin{multline}
L_{CE}(\bm{y},\bm{f}(\bm{x}, \theta))= \sum_{j=1}^N -\bm{y}^j \log(\bm{f}^j) \\
-(1-\bm{y}^j) \log((1-\bm{f}^j))
\end{multline}
Training a segmentation model, therefore, involves finding the model parameters $\theta$ that minimize the adopted loss function on the training data.

%-------------------------------------------------------------------------
\subsection{Ensemble of Segmentation Models}

One can execute the aforementioned training $K$ times using the loss functions defined above to obtain $K$ different parameter solutions $\{\theta_1, \ldots, \theta_K\}$.
Each of these training sessions can rely on slightly different training data as in the case of bagging~\cite{breiman1996bagging} or different random initializations~\cite{uncertainty}.
Given an ensemble of models, the classical approach is to average the individual predictive probabilities, which would be considered as a better approximation of the true posterior:
\begin{equation}
    p_{ensemble}(\bm{y}^j = 1 | \bm{x}) = \frac{1}{K} \sum_{k=1}^K \bm{f}^j(\bm{x}, \theta_k) \label{eq:ensemble}.
\end{equation}
The ensemble probabilistic prediction is usually thresholded with $0.5$ to obtain a binary segmentation. In the remainder of the paper, we refer to this approach as the \textit{baseline ensembling method}.

\subsection{Ensembling Low Precision Models}
In this paper, we propose an ensemble learning strategy that combines diverse models with low precision but high recall. 
Since each model will have a relatively high recall, each model will label the ground truth foreground pixels largely correctly.
On the other hand, there will be many false positives due to the low precision of each model.
If the models are diverse enough, these false positives will largely be very different and thus can cancel out when averaged.

To enforce all models within the ensemble to have high recall, we experimented with two different loss functions: Tversky~\cite{tversky} and balanced cross-entropy loss (BCE)~\cite{BCE}, which are generalizations of the classical Dice and cross-entropy loss functions mentioned above. These loss functions, defined below, have a hyper-parameter that gives the user the control to adopt a different operating point on the precision/recall trade-off. 
\begin{equation}
    \resizebox{\columnwidth}{!}{$L_{Tv}(\bm{y},\bm{f})= 1 - \frac{\sum_{j=1}^N \bm{y}^j\bm{f}^j}{\sum_{j=1}^N \bigg[\bm{y}^j\bm{f}^j + \beta \bm{y}^j(1-\bm{f}^j) + (1-\beta) (1-\bm{y}^j)\bm{f}^j\bigg]}$}
    \label{eq:tversky}
\end{equation}
\begin{multline}
L_{BCE}(\bm{y},\bm{f})= \sum_{j=1}^N -\beta \bm{y}^j \log(\bm{f}^j) \\
- (1-\beta) (1-\bm{y}^j) \log((1-\bm{f}^j)) 
    \label{eq:BCE}
\end{multline}

Note $\beta \in [0,1)$ is a hyper-parameter and plays a similar role for both loss functions. When $\beta=0.5$, Tversky loss becomes equivalent to Dice, and BCE is the same as regular cross-entropy. For higher values of $\beta$, these loss functions will penalize false negatives (pixels with $\bm{y}^j = 1$ and $\bm{f}^j < 0.5$) more than false positives (pixels with $\bm{y}^j = 0$ and $\bm{f}^j > 0.5$). 
E.g., for $\beta > 0.9$, the false negative rates will be kept low (such that the recall rate is greater than $90\%$, for instance) while producing many false positives, yielding low precision.  
One can achieve the opposite effect of high precision but low recall with low values of $\beta$.

The idea we are promoting in this paper is to use the Tversky or BCE loss with a relatively high $\beta$ in training individual models that make up an ensemble. We believe that other loss functions that can be tuned to control the precision/recall trade-off should also work for our purpose. The comparison between using different losses need to be further explored. 
In our experiments, when training each model, we randomly choose a value between $[0.9, 1)$ and use that to define the loss function for that training session in order to promote diversity among individual models. 
The exact range of $\beta$ can be adjusted based on validation performance and desired level of recall.
%Such a high value of $\beta$ can result in low accuracy and low precision, but high recall predictions because of the highly unbalanced penalties on making false negative and false positive errors.  
We then average these predictions in an ensemble, as in Equation~\ref{eq:ensemble}.
We threshold the ensemble prediction at the relatively high value of $0.9$, which we found is effective in removing residual false positives. 
\textcolor{black}{We present results for alternative thresholds in our Supplemental Material. We note that the threshold is an important hyper-parameter that can be tuned for best ensembling performance during validation. In all of our three experiments presented below, however, we simply used a threshold of $0.9$.} 
In an ensemble of ten or fewer models, this strategy is similar to aggregating the individual model segmentation (e.g. obtained after thresholding with 0.5) with a logical AND operation. 

\textcolor{black}{The key of our method is to combine diverse high recall models to reduce the false positives in the final prediction. We empirically observe that a $\beta$ value between $[0.9, 1)$ is sufficient to make the model output to have a recall rate greater than $0.9$. On the other hand, changing the $\beta$ used in the loss function from $0.9$ to $0.99$ effectively changes the relative weights of false positive and false negative from $1:9$ to $1:99$. 
Models trained with different weights of FP and FN have different sets of accessible hypotheses and optimization trajectories in the corresponding hypothesis space \cite{brown2004diversity}. For example, a set of parameters that reduces the false negative rate by $1\%$ at the cost of increasing the false positive rate by $20\%$ might be rejected by models trained with $\beta = 0.9$, but considered as an acceptable hypothesis for models trained with $\beta = 0.99$.
Thus, by randomizing over $\beta$ during training, we can promote more diversity in our ensembles, while achieving a minimum amount of false negative error that can be determined empirically by setting the lower bound in the range of $\beta$.} 

\subsection{Metric for Model Diversity}
To quantify the diversity in an ensemble, we measured the agreement between pairs of models. More specifically, we measured the similarity between two models $\bm{f}_1$ and $\bm{f}_2$ for both true positive and false positive parts of the predictions, and computed the average across all model pairs in an ensemble: 

\begin{equation}
sim(\bm{p}_1,\bm{p}_2) = \frac{2\sum_{j=1}^{N}\bm{p}_1^j\bm{p}_2^j}{\sum_{j=1}^N \bm{p}_1^j + \sum_{j=1}^N \bm{p}_2^j} 
\label{eq:Dice}
\end{equation}

where $\bm{p}^j = \bm{f}^j\times\bm{y}^j$ or $\bm{f}^i\times(1-\bm{y}^j)$ depending on whether we wanted the true positive or false positive similarity. 
Our insight is that in an ideal ensemble, true positive similarity should be high, whereas false positive similarity should be low.
%For the predictions we want to include in the low ensemble model, we want the average similarity of true positive (false positive) to be high (low).  

\section{Experiments}
\subsection{Internal Carotid Artery Lumen Segmentation}
We first implemented our ensemble learning strategy on the challenging task of internal carotid artery (ICA) lumen segmentation. 
Segmentation of the ICA is clinically important because different types of plaque tissue confer different degrees of risk of stroke in patients with carotid atherosclerosis.
Our IRB approved anonymized data-set consists of 76 Computed Tomography Angiogram images collected at [Anonymous Institution]. All CTA images were obtained from patients with unilateral $>50\%$ extracranial carotid artery stenosis. 
All ground truth segmentation labels were created by a human expert [Anonymous Co-author] with clinical experience. 
The ICA lumen pixels were annotated within the volume of $5$ slices above and below the narrowest part of the internal artery, in the vicinity of the carotid artery bifurcation point. 
We first used a neural network-based multi-scale landmark detection algorithm to locate the bifurcation~\cite{bifurcation}, and crop a volumetric patch of $72 \times 72 \times 12$ from the original image with $0.35-0.7$ mm voxel spacing in x and y directions and $0.6-2.5$ mm voxel spacing in the z-direction, preserving the original voxel spacings. 
For each case, we confirmed that the annotated lumen is included in the patch. 
The data were then randomly split into 48 training, 12 validation, and 16 test patients. 
The created 3D image patches were used for all subsequent training and testing. 
We employed a 3D U-Net as our architecture for all of the models we trained in the ensemble, using the same design choices ~\cite{3dunet}. 
In this first experiment, we used the Tversky loss with high $\beta$ values (we also experimented with balanced cross-entropy, and present those results in Supplemental Material). 
We used a mini-batch size of 4, and trained our models using the Adam optimizer~\cite{kingma2014adam} with a learning rate of 0.0001 for 1000 epochs. 

\begin{figure*}
\begin{center}
%\fbox{\rule{0pt}{2in} \rule{.9\linewidth}{0pt}}
\includegraphics[width=0.8\linewidth]{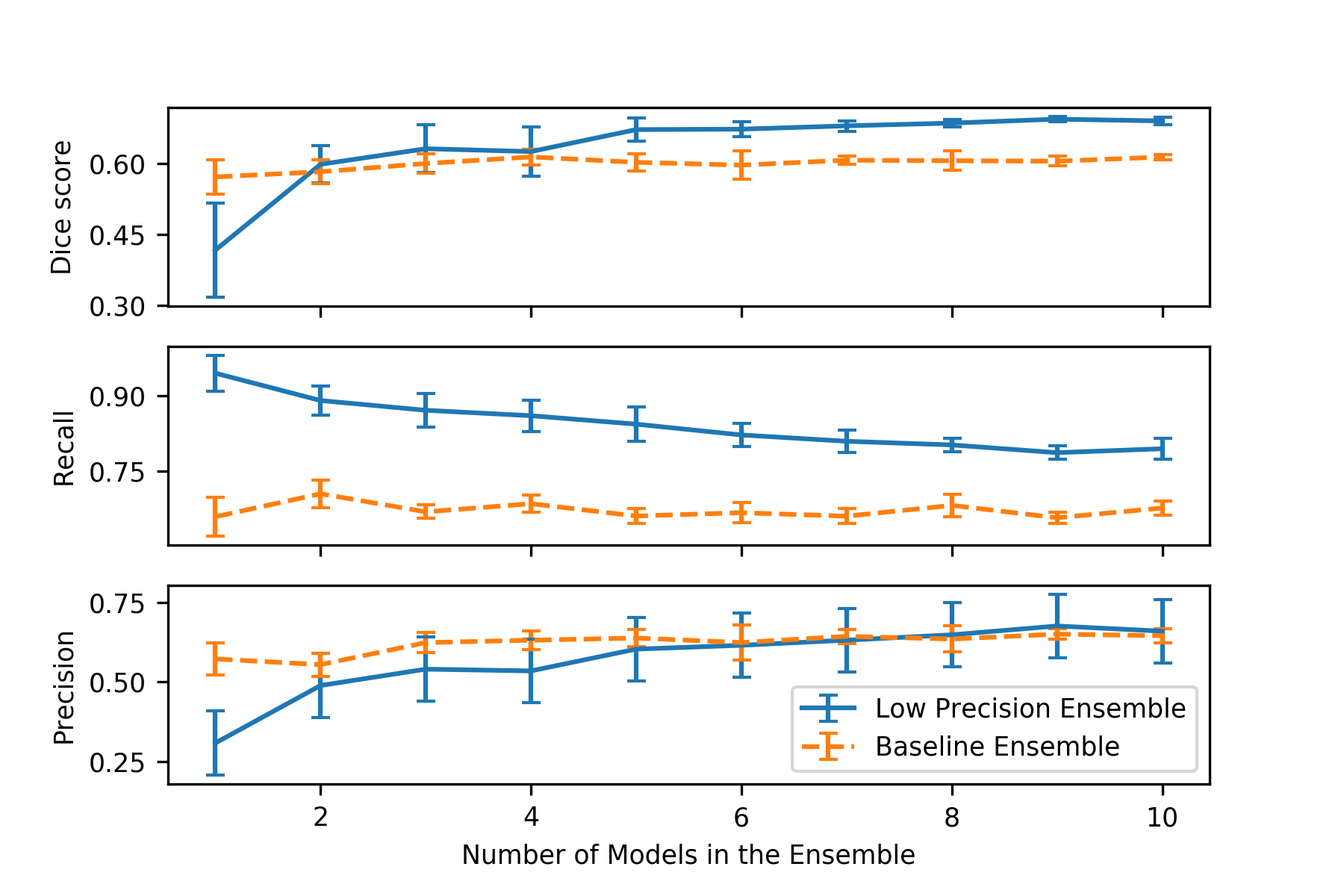}
\end{center}
\caption{Dice score, recall, and precision for two ensemble strategies vs. the number of models in the ensemble for segmentation of internal carotid artery in neck CTA}
\label{fig1}
\end{figure*}

\textcolor{black}{We trained 12 different models with uniform $\beta=0.95$, and another 12 models with uniformly distributed random $\beta$ values ranging from 0.9 to 1, resulting in low precision but high recall models.} All models were randomly initialized, and a distinct set of training and validation images were used to increase model diversity. 
We called them \textit{low prec ensemble ($\beta=0.95$)} and \textit{low prec ensemble (random $\beta$)} respectively.
We also trained another 12 models with $\beta=0.5$, creating a \textit{baseline ensemble} of models trained with Dice loss, random initialization, and random train/validation split. Bagging were used for both baseline and low precision ensemble. 

Figure~\ref{fig1} plots the quantitative results of the two ensemble strategies (low prec ensemble with random $\beta$ and baseline ensemble) for different numbers of models in the ensemble. 
We randomly picked $K$ models from all 12 models to generate the ensemble (where $K$ is between 1 and 10). 
For each $K$, we created 10 random $K$-sized ensembles and computed the mean and standard deviation of the results across the ensembles. 

As the number of models included in the ensemble increases, models trained with regular Dice loss show some improvement in Dice score from 0.576 to 0.614, as well as a small increase in terms of precision. On the other hand, as can be seen for $K=1$ a single low precision model has a relatively low Dice score, but high recall (around $90\%$). The Dice score improves dramatically from 0.435 to 0.708  as we include more low precision models in the ensemble. 
The precision also improves substantially, indicating that many of the false positives are canceling each other. 
The ensemble recall decreases slightly as some of the true positives are removed in the ensemble. 
We observe that in this dataset, the low precision ensemble's performance (Dice) plateaus around 4 models. 

\begin{figure*}
\begin{center}
%\fbox{\rule{0pt}{2in} \rule{.9\linewidth}{0pt}}
\includegraphics[width=0.9\linewidth]{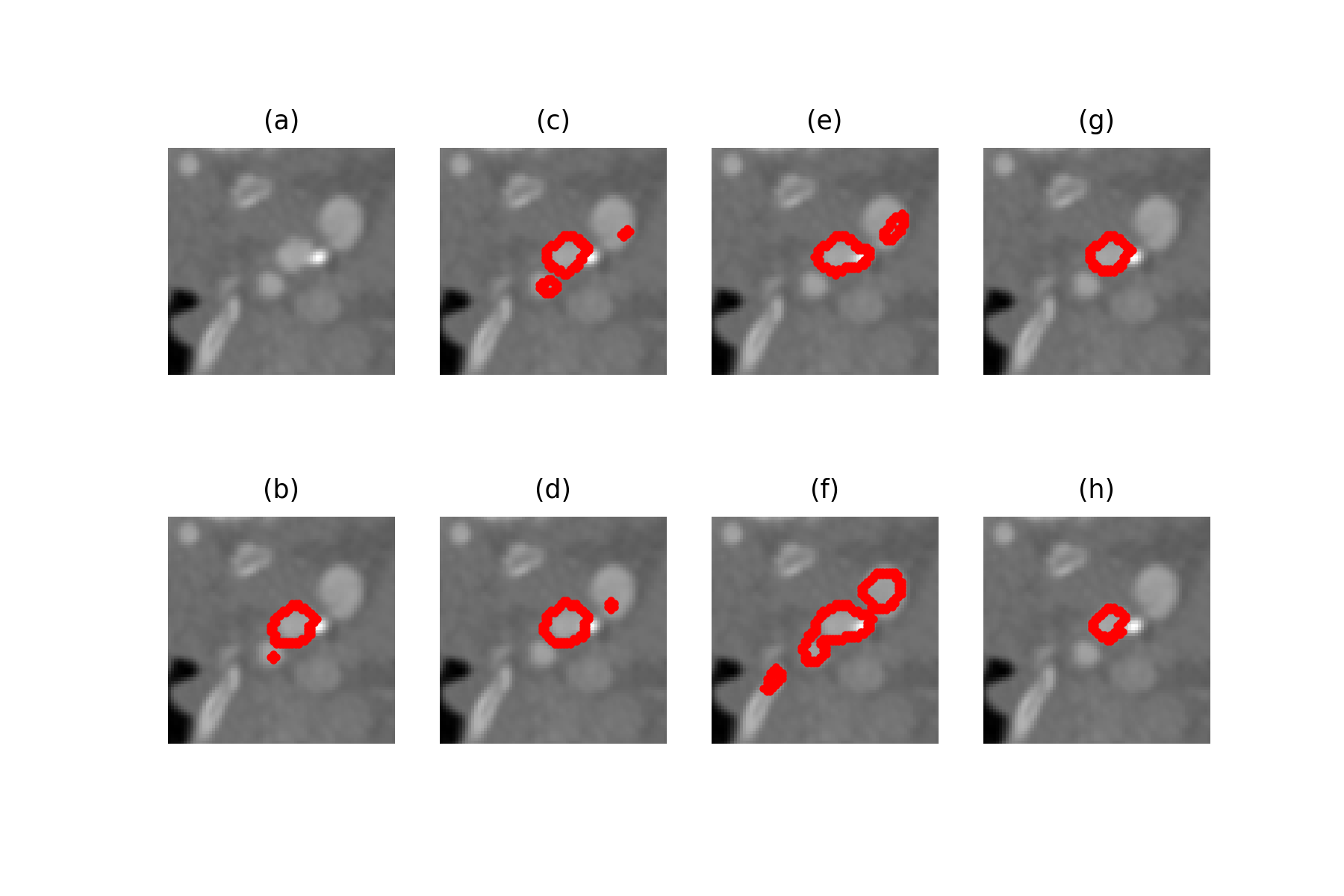}
\end{center}
\caption{An example CTA image and overlaid predictions of the internal carotid artery. (a) is the raw image. (b),(c),(d),(e),(f) are predictions made by different low precision models. (g) is the ensemble result. (h) manual ground-truth}
\label{fig2}
\end{figure*}

Figure~\ref{fig2} visualizes example results from different low precision models trained with random $\beta$. (h) is the ground truth label, and the image has several regions including the internal and external carotid arteries with similar gray values. Note that it is hard even for a human expert to distinguish the internal and external carotid arteries during annotation (see below for inter-human agreement). Fig.~\ref{fig2} (b), (c), (d), (e) and (f) are predictions made by 5 models trained with random $\beta$. We can see that they all make different false positive predictions but capture the structure of interest. (g) is the result after applying our ensemble strategy, and it manages to eliminate most of the false positives.% after applying a relatively high cut-off threshold of $0.9$.   

\begin{table}
\begin{center}
\resizebox{\columnwidth}{!}{
\begin{tabular}{|c|c|c|c|}
\hline
Method &  Dice & Recall & Precision\\
\hline\hline
Single Baseline Model &  $0.576\pm0.302$ & $0.672\pm0.393$ & $0.563\pm0.346$\\
Single Low Precision Model &  $0.435\pm0.192$ & $0.944\pm0.220$ & $0.304\pm0.255$\\
Baseline Ensemble &  $0.614\pm0.294$ & $0.665\pm0.388$ & $0.649\pm0.324$\\
\textcolor{black}{Low prec Ensemble ($\beta$=0.95)} & $0.643\pm0.152$ & $0.736\pm0.191$ & $0.628\pm0.182$\\ 
Low Prec Ensemble (random $\beta$) &  $\bm{0.708}\pm0.170$ & $0.815\pm0.212$ & $0.670\pm0.202$\\
\textcolor{black}{M-Heads~\cite{mhead}} &  $0.655\pm0.134$ & $0.757\pm0.152$ & $0.634\pm0.158$\\
Second Human Expert & $0.791\pm0.140$ & $0.906\pm0.161$ & $0.714\pm0.146$\\
\hline
\end{tabular}
}
\end{center}
\caption{Performance of different methods for Internal Carotid Artery Segmentation in Neck CTA. Best non-manual dice score is \textbf{boldfaced}.}\label{tab1}
\end{table}

Table \ref{tab1} lists the average Dice score, recall, and precision for single baseline and low precision models (random $\beta$s), two ensemble strategies (with $K=12$, both fixed $\beta$ and random $\beta$), \textcolor{black}{an additional top performing ensemble baseline M-Heads~\cite{mhead}}, as well as a secondary manual annotation by another expert who was blinded to the first annotations. 
By using the baseline ensemble strategy with averaging and thresholding at 0.5, we boost the single model dice from 0.576 to 0.614, demonstrating the effectiveness of a regular ensemble strategy in our application. Our low precision ensemble method, on the other hand, is capable of greatly enhancing the performance from a relatively low single model dice of 0.435 to 0.708, utilizing weak segmenters to make a more accurate prediction. \textcolor{black}{We can see that the biggest dice score improvement (from 0.643 to 0.708) comes from using random $\beta$ instead of a single fixed value. The proposed method also has a better dice score compared to the M-Heads method.} 
Compared to the baseline ensemble method, the low-precision ensemble (random $\beta$) has a higher Dice score, a lower false negative rate, and a comparable false positive rate.  
However, there is still room for improvement, particularly in recall rates, as can be observed from the second human expert performance. 

% \begin{figure}[t]
% \begin{center}
% %\fbox{\rule{0pt}{2in} \rule{0.9\linewidth}{0pt}}
% \includegraphics[width=0.8\linewidth]{diversity.png}
% \end{center}
% \caption{Pairwise model similarity of the false positive and true positive predictions in the low precision ensemble.}
% \label{div}
% \end{figure}

\begin{table}
\begin{center}
\resizebox{\columnwidth}{!}{
\begin{tabular}{|c|c|c|c|}
\hline
Method &  True Positive & False Positive & All Positive\\
\hline\hline

Baseline Ensemble &  0.768 & 0.637 & 0.696\\
Low prec Ensemble ($\beta$=0.95) & 0.951 & 0.572 & 0.655\\ 
Low Prec Ensemble (random $\beta$) &  0.954 & 0.515 & 0.643\\

\hline
\end{tabular}
}
\end{center}
\caption{Average pairwise model similarity scores of true positive, false positive, and all positive (foreground) predictions for the different ensemble methods in the neck CTA experiment. Lower values indicate more diversity. A good ensemble should have high diversity in its (e.g. false positive) errors, but less diversity in correct predictions.}
\label{ex1div}
\end{table}

% Figure \ref{div} shows the pairwise similarity among all the low precision models' true positive and false positive predictions. With an average true positive similarity of $0.954$, the correctly identified internal carotid artery is mostly preserved during averaging and thresholding. The mistakes that models made are more diverse among different low precision models, and thus can be canceled during ensemble. 
\textcolor{black}{For our diversity analysis, Table \ref{ex1div} lists the pairwise similarity scores of the true positive, false positive, and all positive (foreground) predictions for different ensemble methods. A higher score means less diverse predictions. Compared to a baseline ensemble, models in the low precision ensemble have a higher (lower) score for their true positive (false positive) predictions. Thus, our low precision ensemble is capable of making more diverse false positive errors but consistent true positive predictions.
We observe that the false positive diversity is higher in the random $\beta$ ensemble, relative to the fixed $\beta=0.95$ ensemble. With an average true positive similarity of $\sim 0.95$, the correctly identified internal carotid artery can be mostly preserved in low precision ensembles.}

%------------------------------------------------------------------------

%-------------------------------------------------------------------------
\subsection{Myocardium Segmentation}
In our second experiment, we employed the dataset from the HVSMR 2016: MICCAI Workshop on Whole-Heart and Great Vessel Segmentation from 3D Cardiovascular MRI~\cite{data}. 
This dataset consists of 10 3D cardiovascular magnetic resonance (CMR) images. 
The image dimension and voxel spacing varied across subjects, and averaged at $390 \times 390 \times 165$ and $0.9 \times 0.9 \times 0.85$ mm. 
The ground truth manual segmentations of both the blood pool and ventricular myocardium are provided. 
In our experiment, we focused on the myocardium segmentation because it is a more challenging task with a low state-of-the-art average dice score.  
Before training, certain pre-processing steps were carried out. 
We first normalized all images to have zero mean and unit variance intensity values. 
Data augmentation was also performed via image rotation and flipping in the axial plane. 
We implemented a 5 fold cross-validation, holding 2 images for testing and 8 images for training. 

To demonstrate that our method is not restricted to a specific network architecture and loss function, we adopted the network and method used by the challenge winner~\cite{fractal}. 
We implemented the 3D FractalNet, with the same parameters and experimental settings proposed by the authors, trained with regular cross-entropy loss. We trained 12 different models, and we call this the \textit{Baseline Ensemble}. 
To train models with high recall and low precision, we used balanced binary cross-entropy loss with random $\beta \in [0.9,1)$. Note that results obtained with Tversky loss are presentede in Supplemental Material. \textcolor{black}{We trained 12 different models to create the low precision ensemble (random $\beta$) and another 12 models with fixed $\beta$=0.95.}

\begin{figure}[t]
\begin{center}
\includegraphics[width=0.8\linewidth]{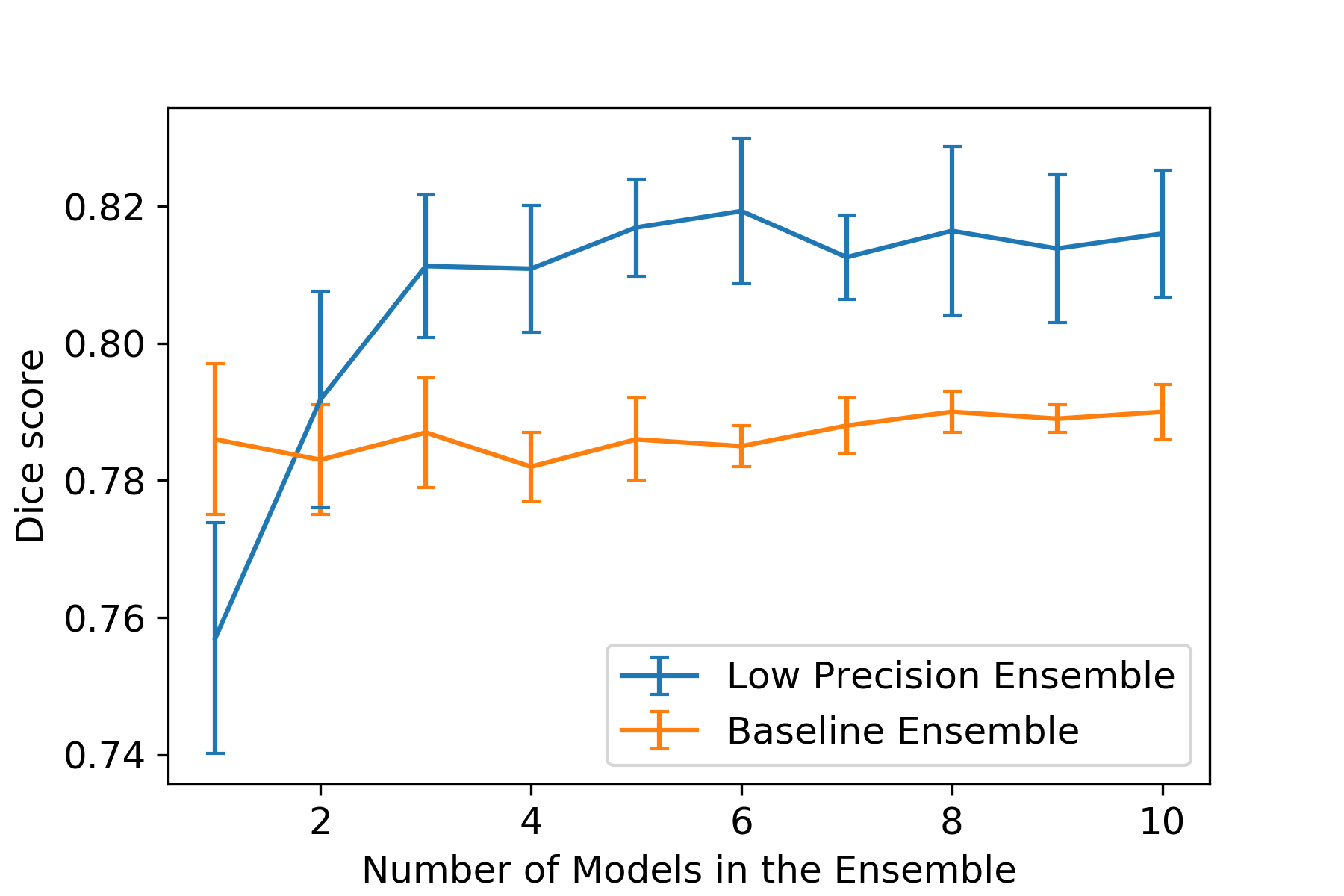}
\end{center}
\caption{Performance of Low-Precision Ensemble vs Number of Models: Segmentation of Ventricular Myocardium in MRI}
\label{fig3}
\end{figure}

\begin{table}
\begin{center}
\resizebox{\columnwidth}{!}{
\begin{tabular}{|c|c|c|c|}
\hline
Method &  Dice & Recall & Precision\\
\hline\hline
Single Baseline Model~\cite{fractal} &  $0.786\pm0.045$ & $0.845\pm0.047$ & $0.747\pm0.081$\\
Single Low Precision Model &  $0.757\pm0.065$ & $0.974\pm0.062$ & $0.607\pm0.121$\\
Baseline Ensemble &  $0.790\pm0.033$ & $0.872\pm0.022$ & $0.750\pm0.078$\\
\textcolor{black}{Low Prec Ensemble ($\beta$=0.95)} &  $0.796\pm0.046$ & $0.904\pm0.028$ & $0.715\pm0.092$\\
Low Prec Ensemble (random $\beta$) &  $\bm{0.815}\pm0.052$ & $0.949\pm0.034$ & $0.719\pm0.097$\\
\hline
\end{tabular}
}
\end{center}
\caption{Performance for Segmentation of Ventricular Myocardium in MRI. Best dice score is \textbf{boldfaced}.}\label{tab2}
\end{table}

%-------------------------------------------------------------------------
Similar to the previous experiment, we randomly picked $K$ models from all 12 models to generate an ensemble, where $K$ goes from 1 to 10. 
For each $K$, we created 10 random $K$-sized ensembles, and computed the mean and standard deviation of the results across the ensembles (see Figure~\ref{fig3}).
Table \ref{tab2} shows the experimental results of the average of 5 fold testing. \textcolor{black}{The low precision ensemble model trained with fixed $\beta=0.95$ value does not show significant improvement over the regular ensemble.} The Dice score improves from 0.790 (for the state-of-the-art baseline) to 0.815 (for the low precision ensemble with random $\beta$). The low precision ensemble with random $\beta$ method has higher Dice score and recall, but lower precision. We also observe a higher improvement in terms of average dice scores from single models to the ensemble. 

\begin{table}
\begin{center}
\resizebox{\columnwidth}{!}{
\begin{tabular}{|c|c|c|c|}
\hline
Method &  True Positive & False Positive & All Positive\\
\hline\hline

Baseline Ensemble &  0.926 & 0.760 & 0.898\\
Low prec Ensemble ($\beta$=0.95) & 0.971 & 0.644 & 0.842\\ 
Low Prec Ensemble (random $\beta$) &  0.974 & 0.621 & 0.818\\

\hline
\end{tabular}
}
\end{center}
\caption{Average pairwise model similarity scores of true positive, false positive, and all positive (foreground) predictions for the different ensemble methods in the myocardium segmentation experiment. Lower values indicate more diversity. A good ensemble should have high diversity in its (e.g. false positive) errors, but less diversity in correct predictions.}
\label{ex2div}
\end{table}

\textcolor{black}{Additionally, we perform a diversity analysis for different ensemble models. As we observe in Table \ref{ex2div}, models in the low precision ensemble have more consistent true positive predictions but more diverse false positive errors.}

\subsection{Multiple Sclerosis (MS) Lesion Segmentation}

\begin{figure*}[!th]
\begin{center}

\subfloat{\includegraphics[width=0.16\textwidth]{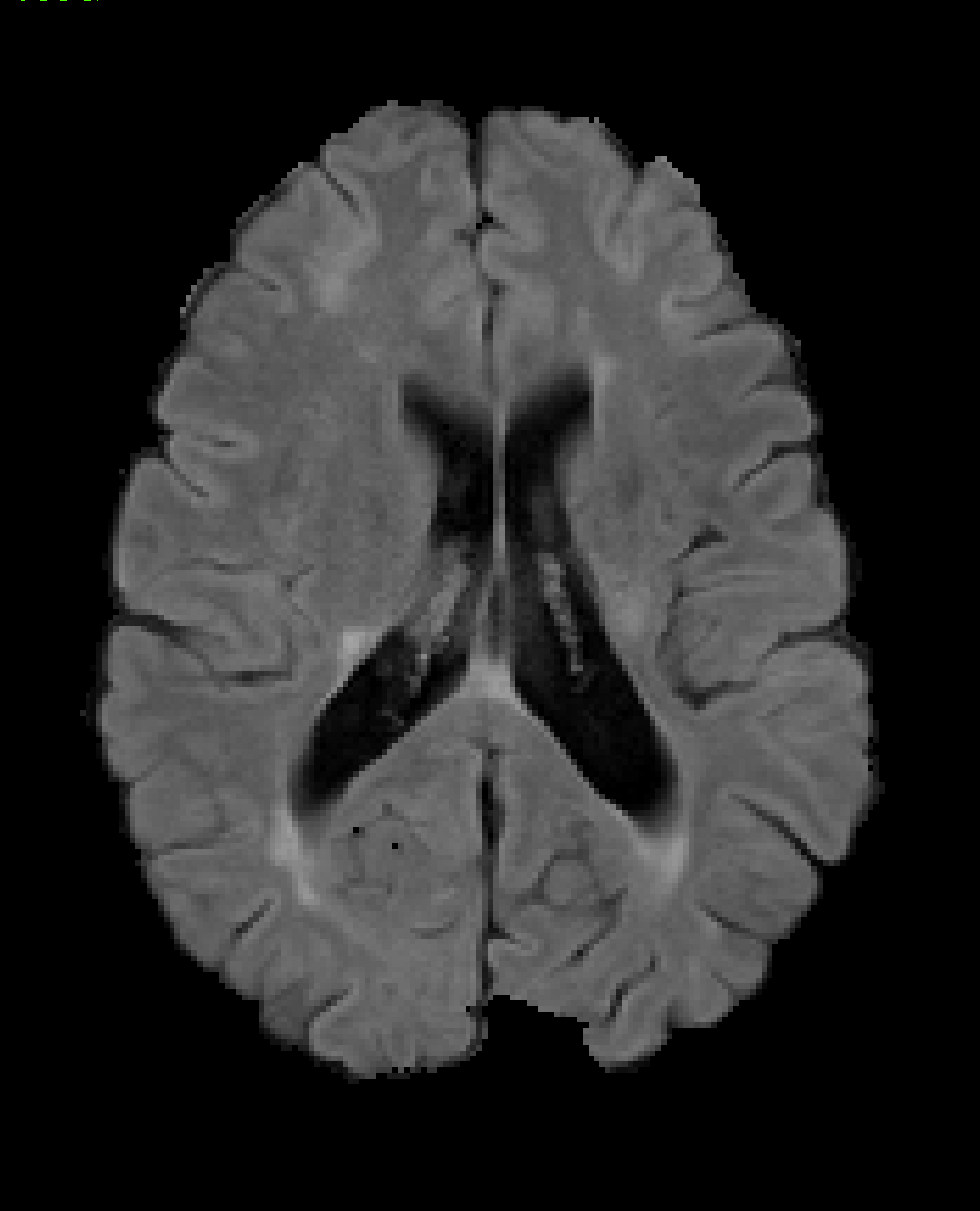}}
\subfloat{\includegraphics[width=0.16\textwidth]{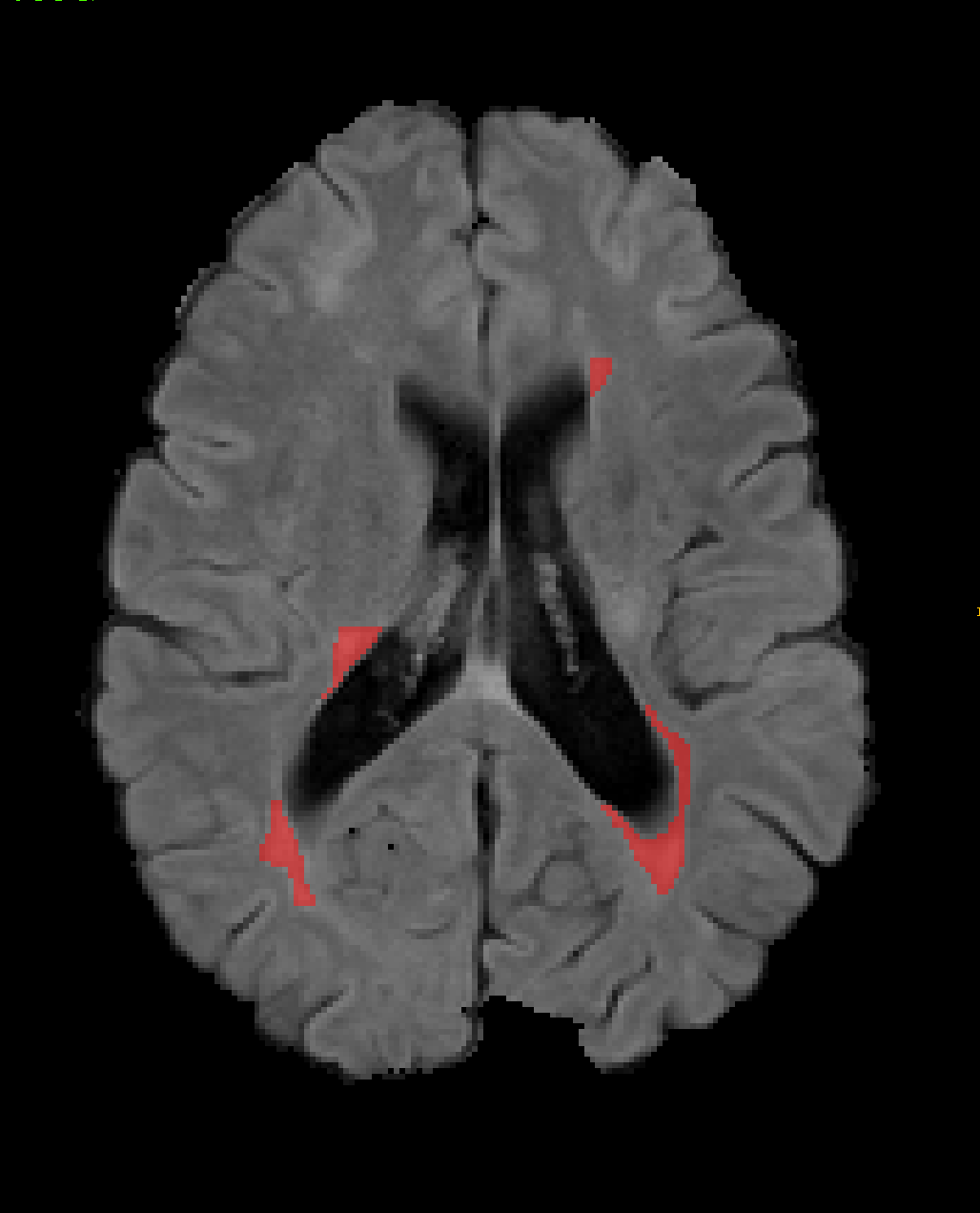}}
\subfloat{\includegraphics[width=0.16\textwidth]{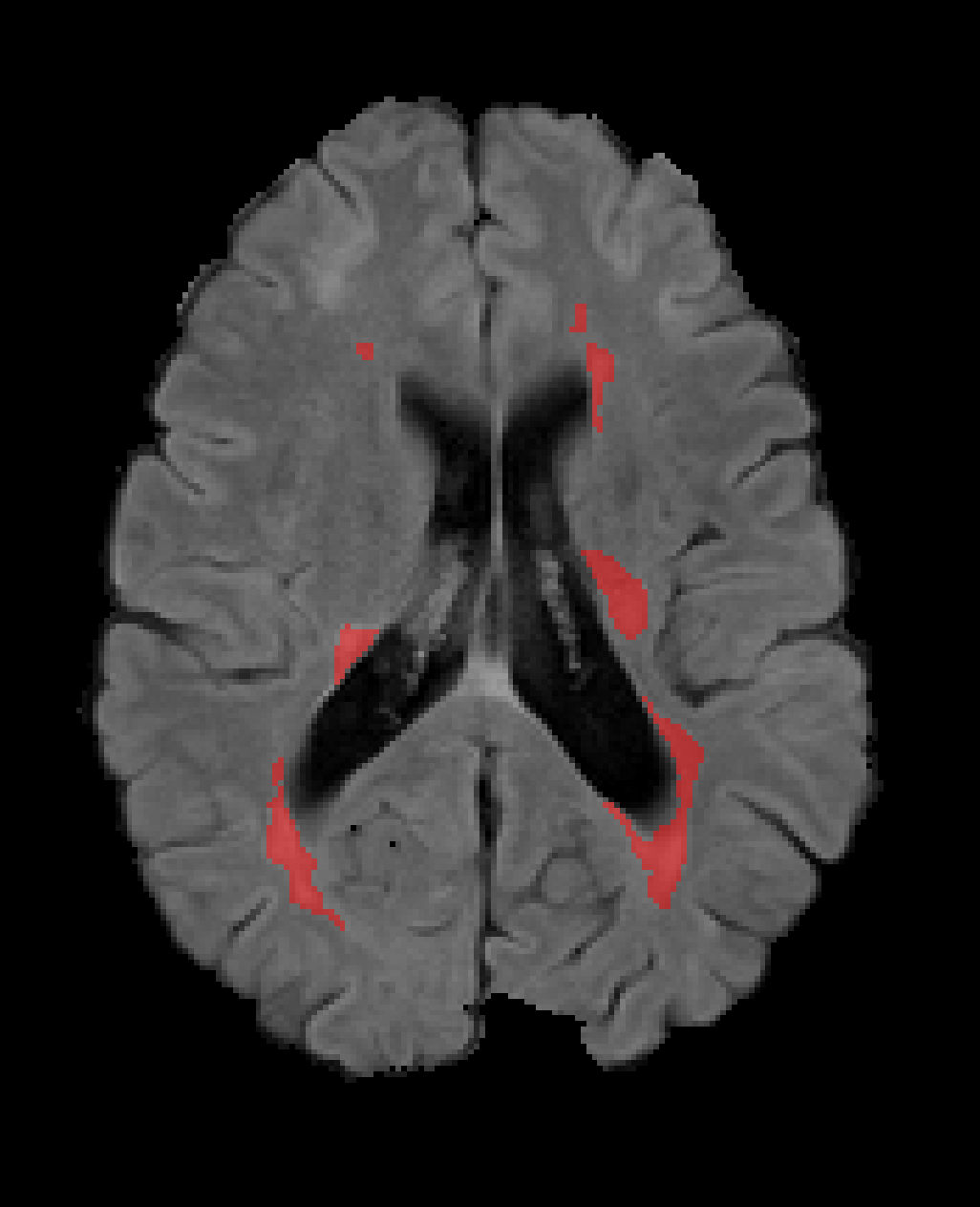}}
\subfloat{\includegraphics[width=0.16\textwidth]{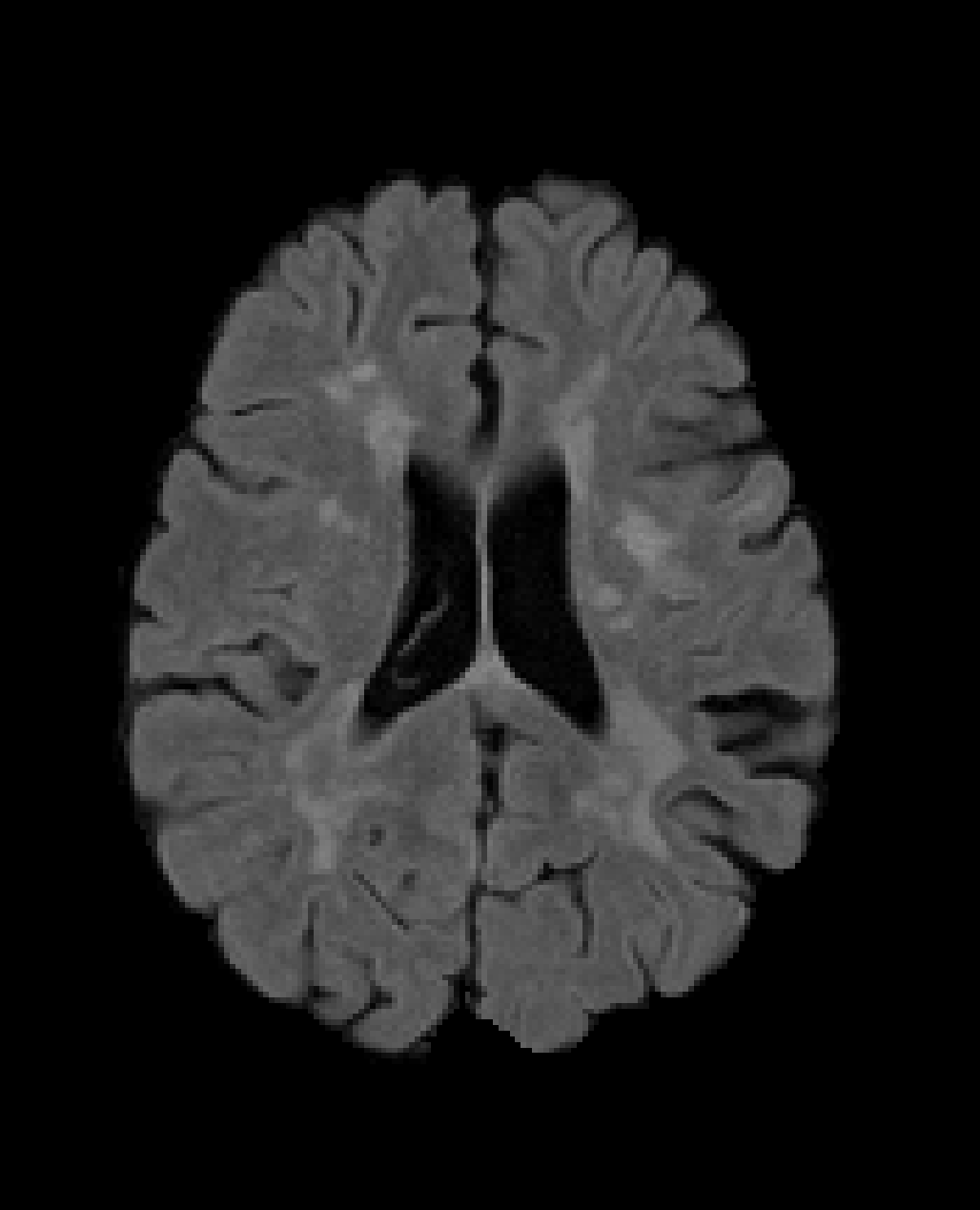}}
\subfloat{\includegraphics[width=0.16\textwidth]{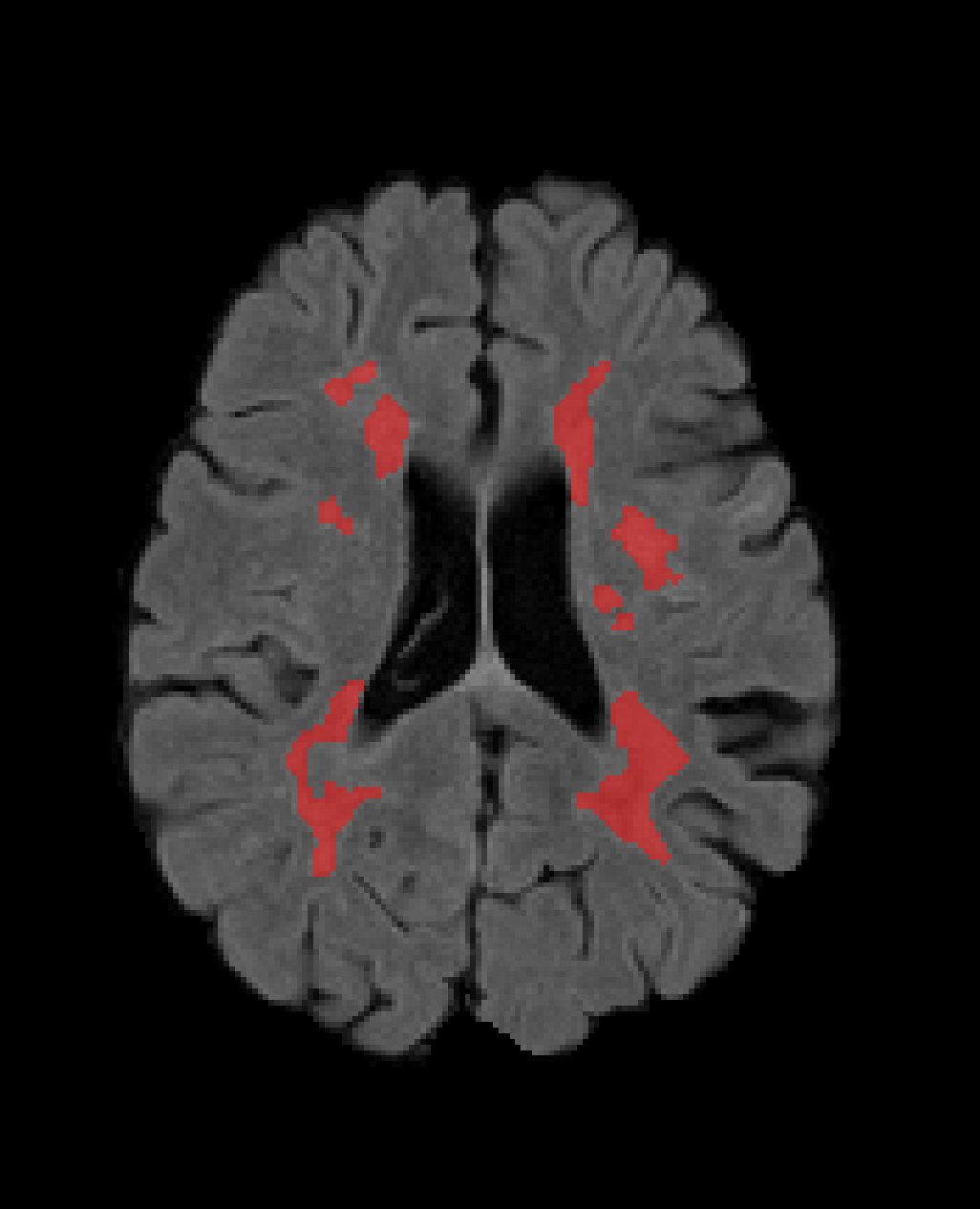}}
\subfloat{\includegraphics[width=0.16\textwidth]{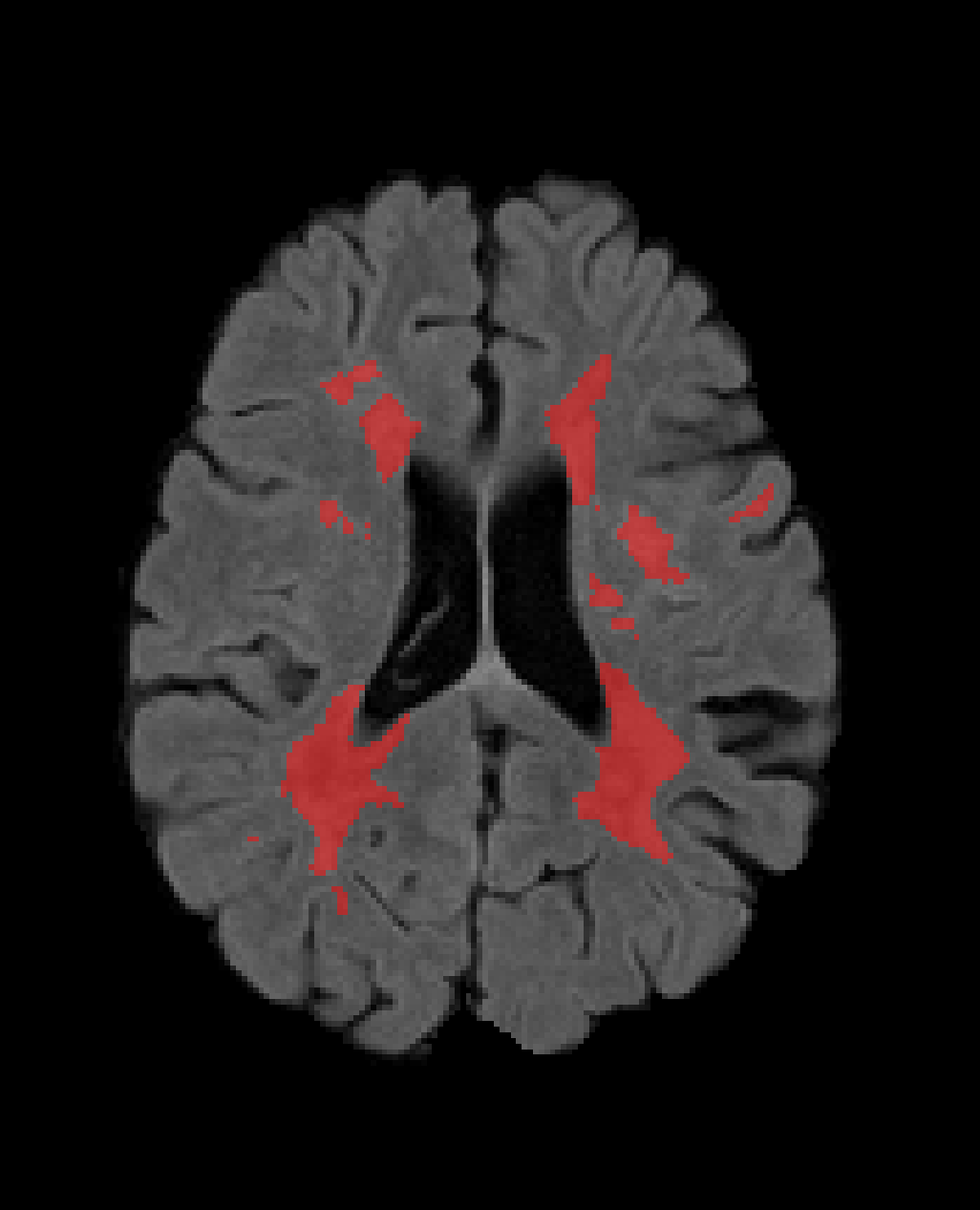}}

\end{center}

\caption{
Example FLAIR images and corresponding masks traced by two human experts and marked in red. 
The left three are a flair image, 1st rater's mask, 2nd rater's mask from subject-01's first time-point scan.
The right three are from subject-02's first time-point scan.
We can see from the figure, besides ms lesion areas, there are many other hyperintensities that can confuse algorithms or even human experts.
}
\label{fig:isbi_samples}
\end{figure*}

We conduct our third experiment on MS lesion segmentation.
MS is a chronic, inflammatory demyelinating disease of the central nervous system in the brain.
Precise segmentation can help characterize MS lesions and provide important markers for clinical diagnosis and disease progress assessment.
However, MS lesion segmentation is challenging and complicated as lesions vary vastly in terms of location, appearance, and shape.
Concurrent hyper-intensities make MS lesion tracing more difficult even for experienced neural radiologists (as shown in Fig~\ref{fig:isbi_samples}).
Dice score between masks traced by two raters from a ISBI dataset is only $0.732$~\cite{carass2017longitudinal}.

We employ the dataset from ISBI 2015 Longitudinal MS Lesion Segmentation Challenge~\cite{carass2017longitudinal} to verify our method.
The ISBI dataset contains MRI scans from 19 subjects, and each subject has 4-6 time-point scans.
For each scan, FLAIR, PD-weighted, T2-weighted, and T1-weighted images are provided.
All image modalities are skull-stripped, dura-stripped, N4 inhomogeneity corrected, and rigidly co-registered to a $1mm$ isotropic MNI template.
Each image contains 182 slices with $FOV = 182\times 256$. 
Two experienced raters manually traced all lesions, so there are two gold-standard masks. 
To our knowledge, using the intersection of the two masks to train our model yields the best performance.
Only 5 training subjects (21 images) have publicly available gold-standard masks.
We can evaluate our model on an online website by submitting predicted results of the remaining 14 subjects (61 images).
Unlike ICA lumen and myocardium segmentation, MS patients usually contain numerous lesions with different sizes.
Thus, to evaluate the performance of the method, we apply metrics of lesion-wise precision (L-precision) and lesion-wise recall (L-recall), which are defined as follows.
\begin{itemize}
    \item Lesion-wise Precision
\end{itemize}
\begin{equation}
    \text{L-Precision} = \frac{LTP}{PL},
\end{equation}
\begin{itemize}
    \item Lesion-wise Recall
\end{itemize}
\begin{equation}
    \text{L-Recall} = \frac{LTP}{GL},
\end{equation}
where $LTP$ denotes the number of lesion-wise true positives, $GL$ is the total number of lesions in the gold-standard segmentation, and $PL$ is the total number of lesions in the predicted segmentation. We calculate lesion-wise accuracy as the average of lesion-wise recall and precision.

\begin{table}
\begin{center}
\resizebox{\columnwidth}{!}{
\begin{tabular}{|c|c|c|}
\hline
Method &  Dice & (L-Recall + L-Precision)/2\\
\hline\hline
Baseline Ensemble~\cite{zhang2019rsanet}    &  $0.624$ & $\frac{0.458+0.889}{2}=0.673$\\
\textcolor{black}{Low Prec Ensemble ($\beta$ = 0.95)}          &  $0.645$ & $\frac{0.473 +0.823}{2}=0.648$\\
Low Prec Ensemble (random $\beta$)           &  $\bm{0.662}$ & $\frac{0.491 + 0.849}{2} = 0.670$\\
MB-Net~\cite{aslani2019multi}               &  $0.611$ & $\frac{0.410 + 0.860}{2} = 0.635$\\
CD-Net~\cite{valverde2017improving}         &  $0.630$ & $\frac{0.367 + 0.847}{2} = 0.607$\\
MS-Net~\cite{ghafoorian2017deep}            &  $0.501$ & $\frac{0.429 + 0.434}{2} = 0.431$\\

\hline
\end{tabular}
}
\end{center}
\caption{Performance comparison for segmenting MS lesions in brain MRI. Best non-manual dice score is \textbf{boldfaced}.}\label{tab3}
\end{table}

We compare our ensemble method with four recent works~\cite{zhang2019rsanet,aslani2019multi,valverde2017improving,ghafoorian2017deep} on MS lesion segmentation. 
We build our baseline network architecture based on a publicly available implementation~\cite{zhang2019rsanet} designed for MS lesion segmentation.
Other three methods for comparison are multi-branch network (MB-Net)~\cite{aslani2019multi}, multi-scale network (MS-Net)~\cite{ghafoorian2017deep}, and cascaded-network (CD-Net)~\cite{valverde2017improving}. 
Similar to myocardium segmentation, all images are normalized to have zero mean and unit variance intensity values.
We use random crop, intensity shifting, and elastic deformation to augment our data.

\textcolor{black}{We trained 10 different models with regular dice loss, 10 models using Tversky loss with random high $\beta \in [0.9,1)$, and another 10 models with high $\beta$ = 0.95 \footnote{Results obtained with balanced cross-entropy loss are included in the Supplemental Material.}. We refer these three methods as \textit{Baseline Ensemble}, \textit{Low Prec Ensemble (random $\beta$)} and \textit{Low Prec Ensemble ($\beta$ = 0.95)} in Table~\ref{tab3}.} 
%All models are averaged together to form a probabilistic map, and thresholds of $0.5$ and $0.9$ are chosen via validation to create the final segmentation map.

We can see from Table~\ref{tab3} that compared with the baseline ensemble model, the low-precision ensemble achieves a lesion-wise accuracy that is approximately the same, but with a lower L-Precision. In terms of overall dice score, the low precision ensemble is $6$ points higher than the baseline ensemble.
Also, compared to the recently proposed MB-Net~\cite{aslani2019multi} , CD-Net~\cite{valverde2017improving}, and MS-Net~\cite{ghafoorian2017deep}, the proposed low precision ensemble exhibits superior performance in all aspects.
%, which shows the superiority of low-precision ensemble in the condition when medical images have different anatomical structures with similar intensity values and unbalanced foreground rate. (see Fig.~\ref{fig:isbi_samples})
We also note that randomizing $\beta$ improves the quality of segmentations, yielding a dice score boost of 1.7 points and an increase in lesion-wise accuracy.

\begin{table}
\begin{center}
\resizebox{\columnwidth}{!}{
\begin{tabular}{|c|c|c|c|}
\hline
Method  & Pairwise Similarity of Foreground Pixels\\
\hline\hline

Baseline Ensemble &  0.814\\
Low prec Ensemble ($\beta$=0.95) & 0.753\\ 
Low Prec Ensemble (random $\beta$) & 0.737\\

\hline
\end{tabular}
}
\end{center}
\caption{Pairwise model similarity scores of foreground predictions for the different ensemble methods. Lower values indicate more diversity.}
\label{ex3div}
\end{table}

\textcolor{black}{Because we do not have ground truth labels for the test images, we cannot show  diversity measurements for true positive and false positive predictions. For overall positive predictions, however, the pairwise similarity scores are listed in Table~\ref{ex3div}. We observe that the low precision ensemble achieves the lowest among all three methods, indicating a more diverse set of results generated by models trained with random high $\beta$'s.}

\section{Conclusion}
In this paper, we presented a novel low-precision ensembling strategy for binary image segmentation. 
Similar to regular ensemble learning, predictions from multiple models are combined by averaging. 

However, in contrast to regular ensemble learning, we encourage the individual models to have a high recall, typically at the expense of low precision and accuracy.
Our goal is to have a diverse ensemble of models that largely capture the foreground pixels, but make different types of false positive predictions that can be canceled after averaging. 

We conducted experiments on three different data-sets, with different loss functions and network architectures. 
The proposed method achieves better Dice score compared to using a single model or a regular ensembling strategy that does not combine high recall models.
We believe that our method can be applied to a wide range of hard segmentation problems, with different loss functions and architectures.
Finally, the proposed strategy can also be used in other types of challenging classification problems, particularly with relatively small classes.

\newpage

{\small
\bibliographystyle{ieee_fullname}
\bibliography{egbib}

\begin{thebibliography}{10}\itemsep=-1pt

\bibitem{aslani2019multi}
Shahab Aslani, Michael Dayan, Loredana Storelli, Massimo Filippi, Vittorio
  Murino, Maria~A Rocca, and Diego Sona.
\newblock Multi-branch convolutional neural network for multiple sclerosis
  lesion segmentation.
\newblock {\em NeuroImage}, 196:1--15, 2019.

\bibitem{breiman1996bagging}
Leo Breiman.
\newblock Bagging predictors.
\newblock {\em Machine learning}, 24(2):123--140, 1996.

\bibitem{brown2004diversity}
Gavin Brown.
\newblock {\em Diversity in neural network ensembles}.
\newblock PhD thesis, Citeseer, 2004.

\bibitem{lung}
Peng Cao, Jinzhu Yang, Wei Li, Dazhe Zhao, and Osmar Zaiane.
\newblock Ensemble-based hybrid probabilistic sampling for imbalanced data
  learning in lung nodule cad.
\newblock {\em Computerized Medical Imaging and Graphics}, 38(3):137--150,
  2014.

\bibitem{carass2017longitudinal}
Aaron Carass, Snehashis Roy, Amod Jog, Jennifer~L Cuzzocreo, Elizabeth Magrath,
  Adrian Gherman, Julia Button, James Nguyen, Ferran Prados, Carole~H Sudre,
  et~al.
\newblock Longitudinal multiple sclerosis lesion segmentation: resource and
  challenge.
\newblock {\em NeuroImage}, 148:77--102, 2017.

\bibitem{3dunet}
{\"O}zg{\"u}n {\c{C}}i{\c{c}}ek, Ahmed Abdulkadir, Soeren~S Lienkamp, Thomas
  Brox, and Olaf Ronneberger.
\newblock 3d u-net: learning dense volumetric segmentation from sparse
  annotation.
\newblock In {\em International conference on medical image computing and
  computer-assisted intervention}, pages 424--432. Springer, 2016.

\bibitem{ghafoorian2017deep}
Mohsen Ghafoorian, Nico Karssemeijer, Tom Heskes, Mayra Bergkamp, Joost
  Wissink, Jiri Obels, Karlijn Keizer, Frank-Erik de Leeuw, Bram van Ginneken,
  Elena Marchiori, et~al.
\newblock Deep multi-scale location-aware 3d convolutional neural networks for
  automated detection of lacunes of presumed vascular origin.
\newblock {\em NeuroImage: Clinical}, 14:391--399, 2017.

\bibitem{kamnitsas2017ensembles}
Konstantinos Kamnitsas, Wenjia Bai, Enzo Ferrante, Steven McDonagh, Matthew
  Sinclair, Nick Pawlowski, Martin Rajchl, Matthew Lee, Bernhard Kainz, Daniel
  Rueckert, et~al.
\newblock Ensembles of multiple models and architectures for robust brain
  tumour segmentation.
\newblock In {\em International MICCAI Brainlesion Workshop}, pages 450--462.
  Springer, 2017.

\bibitem{threshold}
Anuj Karpatne, Ankush Khandelwal, and Vipin Kumar.
\newblock Ensemble learning methods for binary classification with
  multi-modality within the classes.
\newblock In {\em Proceedings of the 2015 SIAM International Conference on Data
  Mining}, pages 730--738. SIAM, 2015.

\bibitem{kingma2014adam}
Diederik~P Kingma and Jimmy Ba.
\newblock Adam: A method for stochastic optimization.
\newblock {\em arXiv preprint arXiv:1412.6980}, 2014.

\bibitem{krogh1995neural}
Anders Krogh and Jesper Vedelsby.
\newblock Neural network ensembles, cross validation, and active learning.
\newblock In {\em Advances in neural information processing systems}, pages
  231--238, 1995.

\bibitem{uncertainty}
Balaji Lakshminarayanan, Alexander Pritzel, and Charles Blundell.
\newblock Simple and scalable predictive uncertainty estimation using deep
  ensembles.
\newblock In {\em Advances in neural information processing systems}, pages
  6402--6413, 2017.

\bibitem{logicAND}
Chaofeng Li, Guoce Zhu, Xiaojun Wu, and Yuanquan Wang.
\newblock False-positive reduction on lung nodules detection in chest
  radiographs by ensemble of convolutional neural networks.
\newblock {\em IEEE Access}, 6:16060--16067, 2018.

\bibitem{li2018fully}
Hongwei Li, Gongfa Jiang, Jianguo Zhang, Ruixuan Wang, Zhaolei Wang, Wei-Shi
  Zheng, and Bjoern Menze.
\newblock Fully convolutional network ensembles for white matter
  hyperintensities segmentation in mr images.
\newblock {\em NeuroImage}, 183:650--665, 2018.

\bibitem{bifurcation}
Tianyu Ma, Ajay Gupta, and Mert~R Sabuncu.
\newblock Volumetric landmark detection with a multi-scale shift equivariant
  neural network.
\newblock In {\em 2020 IEEE 17th International Symposium on Biomedical Imaging
  (ISBI)}, pages 981--985. IEEE, 2020.

\bibitem{retina}
Debapriya Maji, Anirban Santara, Pabitra Mitra, and Debdoot Sheet.
\newblock Ensemble of deep convolutional neural networks for learning to detect
  retinal vessels in fundus images.
\newblock {\em arXiv preprint arXiv:1603.04833}, 2016.

\bibitem{data}
Danielle~F Pace, Adrian~V Dalca, Tal Geva, Andrew~J Powell, Mehdi~H Moghari,
  and Polina Golland.
\newblock Interactive whole-heart segmentation in congenital heart disease.
\newblock In {\em International Conference on Medical Image Computing and
  Computer-Assisted Intervention}, pages 80--88. Springer, 2015.

\bibitem{park2005connectivity}
Jong~Won Park et~al.
\newblock Connectivity-based local adaptive thresholding for carotid artery
  segmentation using mra images.
\newblock {\em Image and Vision Computing}, 23(14):1277--1287, 2005.

\bibitem{ensemble}
Robi Polikar.
\newblock Ensemble learning.
\newblock In {\em Ensemble machine learning}, pages 1--34. Springer, 2012.

\bibitem{unet}
Olaf Ronneberger, Philipp Fischer, and Thomas Brox.
\newblock U-net: Convolutional networks for biomedical image segmentation.
\newblock In {\em International Conference on Medical image computing and
  computer-assisted intervention}, pages 234--241. Springer, 2015.

\bibitem{mhead}
Christian Rupprecht, Iro Laina, Robert DiPietro, Maximilian Baust, Federico
  Tombari, Nassir Navab, and Gregory~D Hager.
\newblock Learning in an uncertain world: Representing ambiguity through
  multiple hypotheses.
\newblock In {\em Proceedings of the IEEE International Conference on Computer
  Vision}, pages 3591--3600, 2017.

\bibitem{tversky}
Seyed Sadegh~Mohseni Salehi, Deniz Erdogmus, and Ali Gholipour.
\newblock Tversky loss function for image segmentation using 3d fully
  convolutional deep networks.
\newblock In {\em International Workshop on Machine Learning in Medical
  Imaging}, pages 379--387. Springer, 2017.

\bibitem{automate}
Neeraj Sharma and Lalit~M Aggarwal.
\newblock Automated medical image segmentation techniques.
\newblock {\em Journal of medical physics/Association of Medical Physicists of
  India}, 35(1):3, 2010.

\bibitem{pulmonary}
Atsushi Teramoto, Hiroshi Fujita, Osamu Yamamuro, and Tsuneo Tamaki.
\newblock Automated detection of pulmonary nodules in pet/ct images: Ensemble
  false-positive reduction using a convolutional neural network technique.
\newblock {\em Medical physics}, 43(6Part1):2821--2827, 2016.

\bibitem{valverde2017improving}
Sergi Valverde, Mariano Cabezas, Eloy Roura, Sandra Gonz{\'a}lez-Vill{\`a},
  Deborah Pareto, Joan~C Vilanova, Llu{\'\i}s Rami{\'o}-Torrent{\`a}, {\`A}lex
  Rovira, Arnau Oliver, and Xavier Llad{\'o}.
\newblock Improving automated multiple sclerosis lesion segmentation with a
  cascaded 3d convolutional neural network approach.
\newblock {\em NeuroImage}, 155:159--168, 2017.

\bibitem{wang2015hierarchical}
Shuangling Wang, Yilong Yin, Guibao Cao, Benzheng Wei, Yuanjie Zheng, and
  Gongping Yang.
\newblock Hierarchical retinal blood vessel segmentation based on feature and
  ensemble learning.
\newblock {\em Neurocomputing}, 149:708--717, 2015.

\bibitem{BCE}
Saining Xie and Zhuowen Tu.
\newblock Holistically-nested edge detection.
\newblock In {\em Proceedings of the IEEE international conference on computer
  vision}, pages 1395--1403, 2015.

\bibitem{fractal}
Lequan Yu, Xin Yang, Jing Qin, and Pheng-Ann Heng.
\newblock 3d fractalnet: dense volumetric segmentation for cardiovascular mri
  volumes.
\newblock In {\em Reconstruction, segmentation, and analysis of medical
  images}, pages 103--110. Springer, 2016.

\bibitem{zhang2019rsanet}
Hang Zhang, Jinwei Zhang, Qihao Zhang, Jeremy Kim, Shun Zhang, Susan~A
  Gauthier, Pascal Spincemaille, Thanh~D Nguyen, Mert Sabuncu, and Yi Wang.
\newblock Rsanet: Recurrent slice-wise attention network for multiple sclerosis
  lesion segmentation.
\newblock In {\em International Conference on Medical Image Computing and
  Computer-Assisted Intervention}, pages 411--419. Springer, 2019.

\bibitem{zhang2019confidence}
Zhilu Zhang, Adrian~V Dalca, and Mert~R Sabuncu.
\newblock Confidence calibration for convolutional neural networks using
  structured dropout.
\newblock {\em arXiv preprint arXiv:1906.09551}, 2019.

\end{thebibliography}
}

\end{document}